\documentclass[aps,prl,twocolumn,groupedaddress,showpacs]{revtex4-1}
\pdfoutput=1
\usepackage[utf8]{inputenc}
\usepackage[english]{babel}
\usepackage{amsmath,graphicx,enumerate}
\usepackage{epstopdf}
\usepackage{graphicx}
\usepackage{amssymb,amsfonts,amsmath,longtable}
\usepackage{bigints}
\usepackage{slashbox}
\usepackage{pict2e}

\usepackage{multirow}

\usepackage{color}

\begin{document}

\title{Dynamical collective memory in fluidized granular materials} 
\author{A. Plati$^{1}$}
\author{A. Baldassarri$^{2}$}
\author{A. Gnoli$^{2}$}
\author{G. Gradenigo$^{3}$}
\author{A. Puglisi$^{2}$} 
\affiliation{$^1$Dipartimento di Fisica, Universit\`a di Roma
  Sapienza, P.le Aldo Moro 2, 00185, Rome, Italy \\ $^2$Istituto dei
  Sistemi Complessi - CNR and Dipartimento di Fisica, Universit\`a di
  Roma Sapienza, P.le Aldo Moro 2, 00185, Rome, Italy \\ $^3$NANOTEC -
  CNR and Dipartimento di Fisica, Universit\`a di Roma Sapienza, P.le
  Aldo Moro 2, 00185, Rome, Italy }

\date{\today}

\begin{abstract}
Recent experiments with rotational diffusion of a probe in a vibrated
granular media revealed a rich scenario, ranging from the dilute gas
to the dense liquid with cage effects and an unexpected superdiffusive
behavior at large times. Here we setup a simulation that reproduces
quantitatively the experimental observations and allows us to
investigate the properties of the host granular medium, a task not
feasible in the experiment. We discover a persistent collective
rotational mode which emerges at high density and low granular
temperature: a macroscopic fraction of the medium slowly rotates,
randomly switching direction after very long times. Such a rotational
mode of the host medium is the origin of probe's superdiffusion.
Collective motion is accompanied by a kind of dynamical heterogeneity
at intermediate times (in the cage stage) followed by a strong
reduction of fluctuations at late times, when superdiffusion sets in.
\end{abstract}


\maketitle

{\em Introduction} - Granular media are systems made of
macroscopic particles, shortened ``grains'', whose diameter usually
exceeds the hundreds of microns, ideally without an upper limit~\cite{jaeger96b,andreotti13}. A
typical list of granular materials includes sand, powders, cereals,
cements, pharmaceuticals, but in certain contexts may incorporate
planetary rocks such those in Saturn's rings~\cite{brilliantov2015size}. The study of granular
media originates from the several applications in chemistry, material
sciences and in the management of geophysical hazards. Since several
decades, however, it has become a fertile inspiration for theoretical
physics, particularly in the framework of non-equilibrium statistical
mechanics~\cite{poschel03,abate2008effective,ness2018shaken}. In fact, grains can be described as particles
which interact dissipatively~\cite{poeschel}. Even in dilute conditions and under
strong vibro-fluidization, the resemblance between a granular gas and
a molecular gas is deceptive, hiding important dynamical differences~\cite{puglio15}. When
packing fraction increases or external energy input decreases, the
dissipative nature of granular interactions becomes more and more
relevant, making the paradigms of equilibrium statistical physics
inapplicable~\cite{forterre2008flows}.

A state of driven granular matter which has received less attention
than others is the liquid one, i.e. a regime of steady agitation,
loosely interpreted as ``ergodic'', where the spatial arrangement of
grains is not a crystal but still shows some degree of order as in molecular
liquids~\cite{puglisi12,kou2017}. Granular liquids, because of dissipative
interactions, display spatial correlations in the
velocity field, a property which leads to strong violations of the
Fluctuation-Dissipation relation and to interesting memory
effects~\cite{puglisi2007violation,sarra10b,GPSV14}. Another common granular feature is the lack of
energy equipartition, for instance in the inhomogeneity of
``granular temperature'' measured along different directions in an
anisotropic setup~\cite{feitosa2002breakdown}.

\begin{figure}[ht!]
  \parbox{0.48\columnwidth}{
    \raggedleft
    \includegraphics[width=0.4\columnwidth]{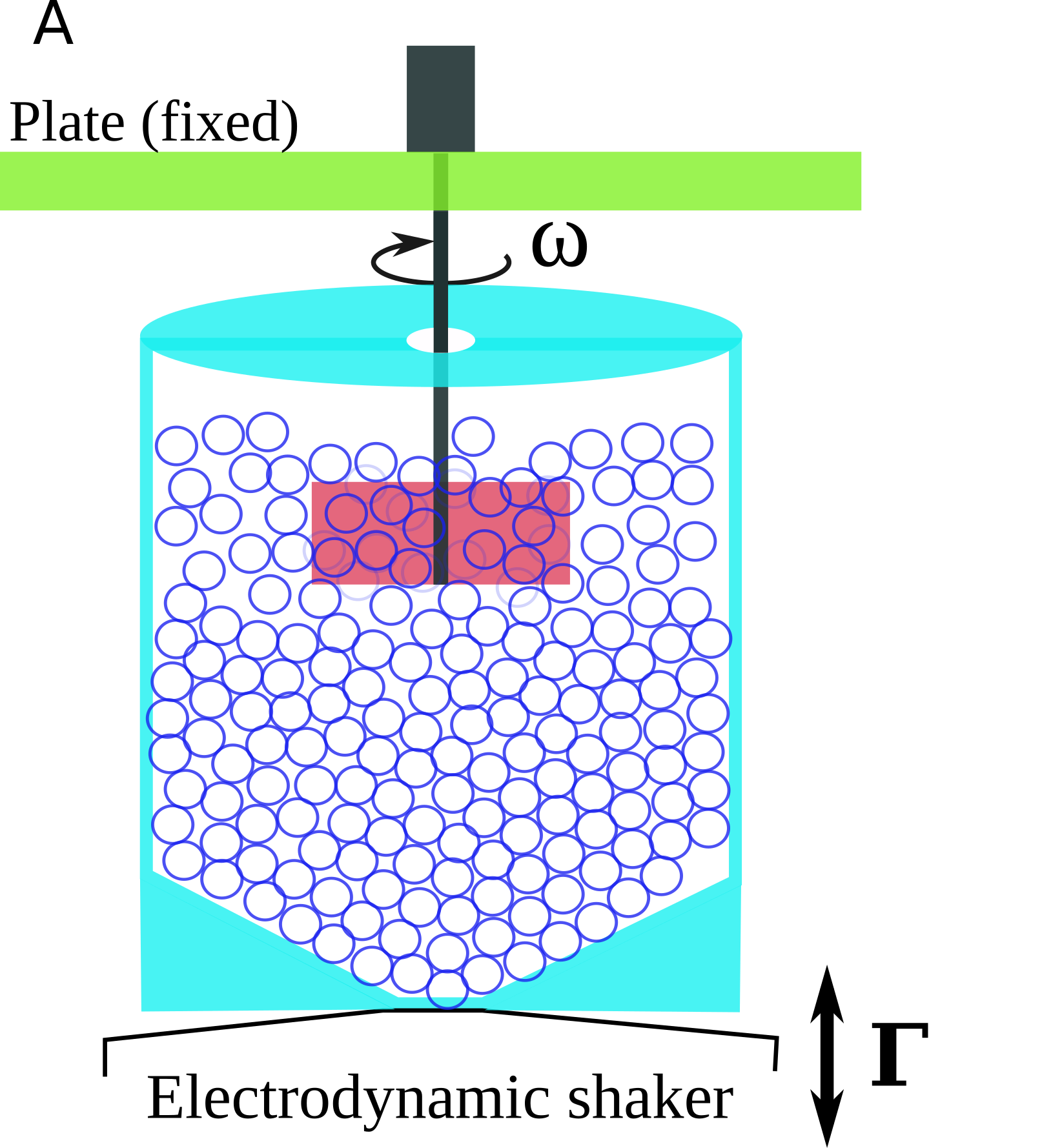}
  }
  \parbox{0.48\columnwidth}{  \includegraphics[width=0.48\columnwidth]{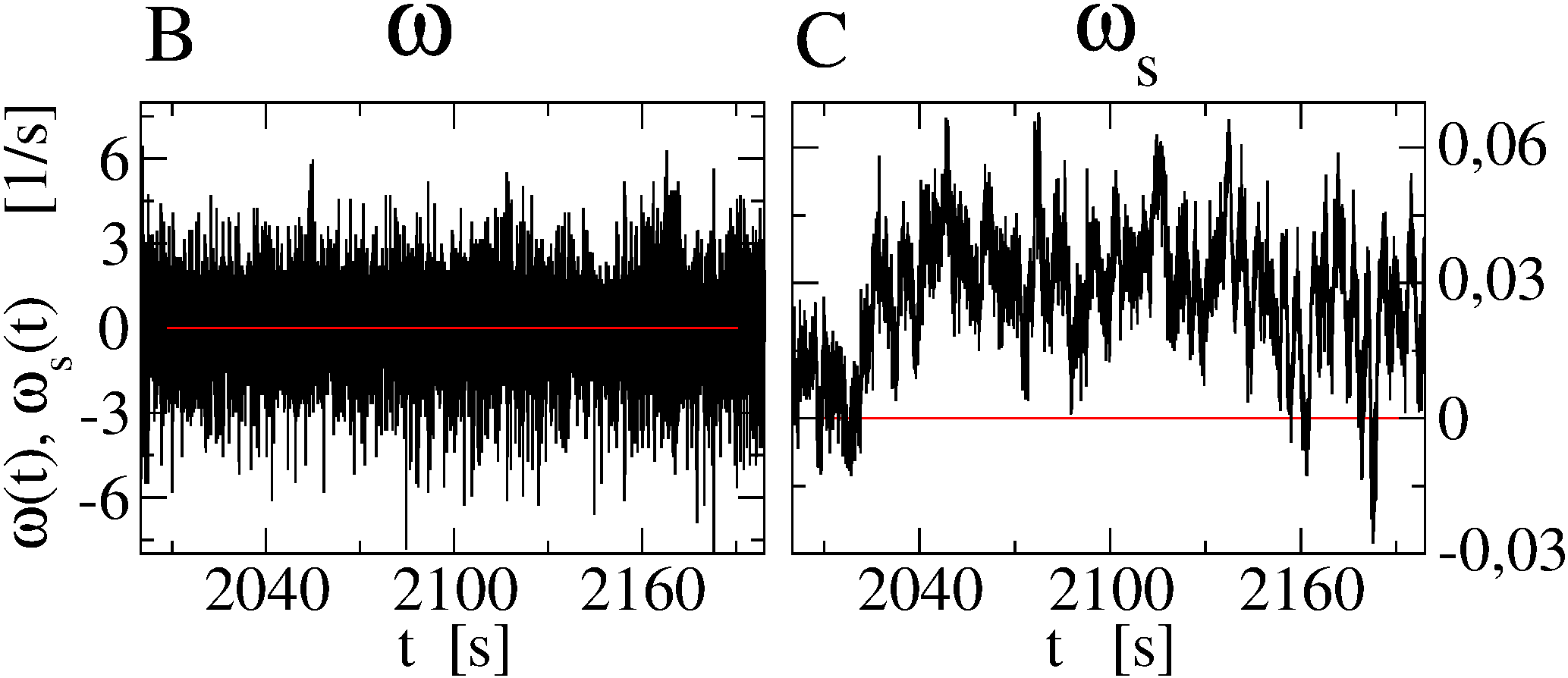}\\ \includegraphics[width=0.48\columnwidth]{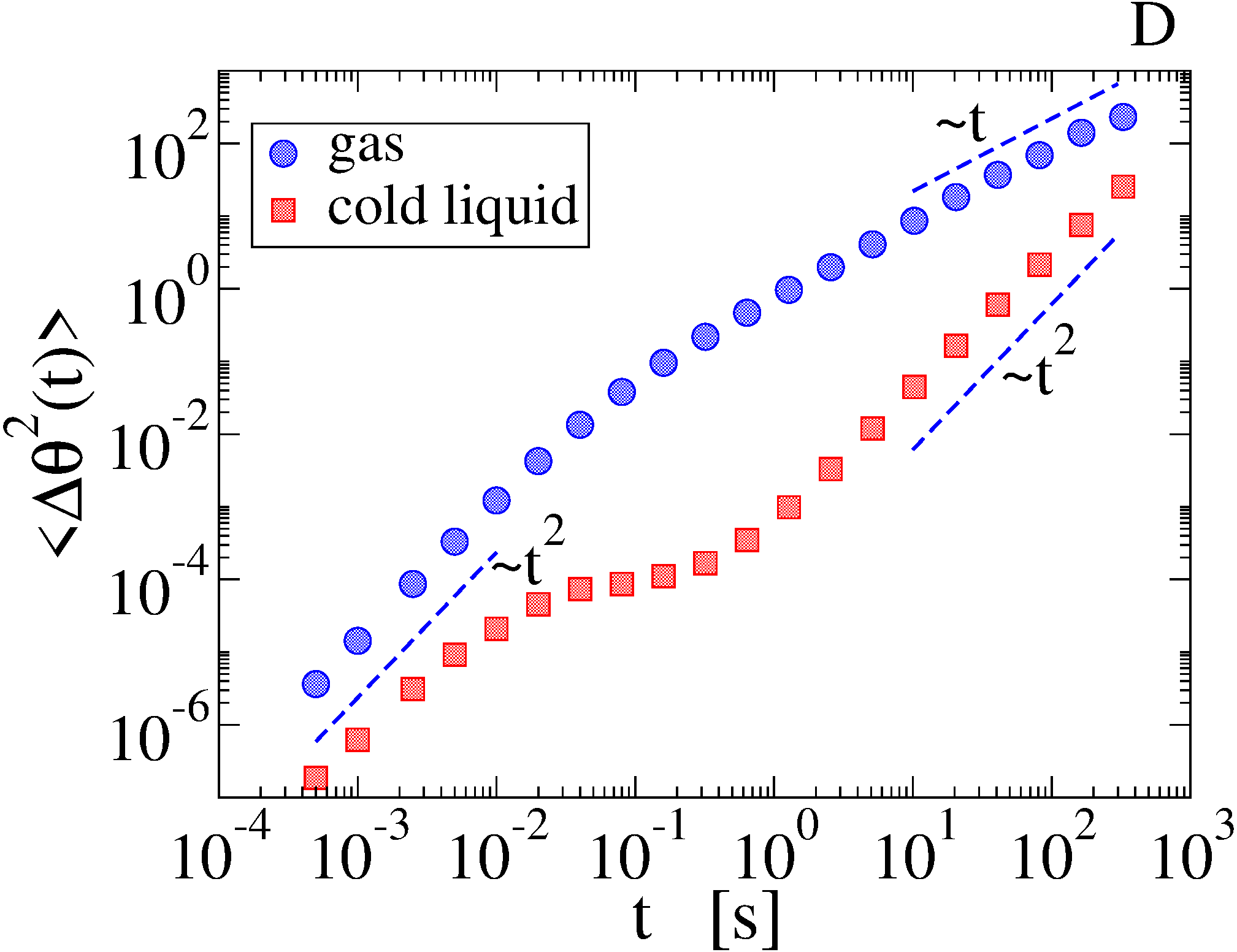}  }
  \caption{A: setup of the experiment and of the
    simulation. B and C: a sample of the $\omega(t)$ signal (left)
    and its filtered ``slow-component''
    $\omega_s=(1/\tau)\int_t^{t+\tau}\omega(t')dt'$ (with $\tau=2$
    seconds) when $N=2600$ and
    $\Gamma=26.8$. D: msd in a
    gas-like case ($N=350$ and $\Gamma=39.8$) and in the cold-liquid
    case ($N=2600$ and $\Gamma=26.8$) with both cage effects and
    late-time superdiffusion.  \label{fig:setup}}
\end{figure}

Recently, some of us have conducted a series of experiments with
vibrofluidized granular materials, using a blade rotating around a
suspended vertical axis as a sort of granular Brownian probe (see
Fig.~\ref{fig:setup}A)~\cite{scalliet}. When the occupied volume fraction is low and
vibration is strong, the medium behaves as a gas and the probe
displays the typical ballistic-diffusive behavior in the angular mean squared
displacement (msd, see Fig.~\ref{fig:setup}D). At high density and weak
vibration, the medium becomes a cold liquid and the msd reveals cage
effects typical of attempted dynamical arrest. Most importantly,
superdiffusion is observed at time delays larger than the onset
of the transient cage effect. Filtering out high frequencies of the
probe's angular velocity, one sees a surprisingly long
dynamical memory, which may induce quasi-ballistic (angular) flights
of the probe for times of the order of tens of seconds
(see~\ref{fig:setup}B and~\ref{fig:setup}C). Up to now only phenomenological models have
caught this behavior~\cite{lasanta2015itinerant}, leaving unexplored its microscopic origin, which
certainly resides in an unconventional dynamics of the granular host
medium. Indeed, a crucial element of those simplified models is an
emergent, effective, giant momentum of inertia of the probe, which may
compare with that of the whole granular
system. Analogies may be found in recent experimental works, where
persistent collective motion~\cite{briand2018,kou2018} or anomalous
diffusion~\cite{kou2017} have been observed in vibrofluidized granular
materials made of anisotropic particles, a crucial difference
with the work discussed here.  In this Letter we reproduce the
experimental observations for the probe through a 3D discrete-elements
model of the real setup including grains' rotations and dissipation
through normal and tangential interactions. After having shown the
quantitative agreement between simulations and the experiment, we
focus on the granular medium alone. Our results give
direct evidence of a collective dynamical memory, originated in a kind
of synchronization between grains movements over long time-scales.


{\em Experimental and numerical setup.} The experiment, realized
in~\cite{scalliet}, is sketched in Fig.~\ref{fig:setup}A. Here we
report the results of a numerical simulation which is meant to
reproduce the setup (container, grains and rotator) in its spatial and
temporal proportions. Both the real and the simulated systems consist
of a cylinder-shaped recipient (diameter 90 mm, maximum height 47 mm
and total volume 245 $\textrm{cm}^3$), with an inverse-conical-shaped
base, containing a number $N$ of steel spheres (diameter $4$ mm,
mass $0.27$ g), representing the ``granular medium''. The recipient is
vertically vibrated: the real system was mounted on an electrodynamic
shaker fed with a noisy signal (spectrum approximately flat in the
range $200-400$ Hz and roughly empty outside that range), in the
simulated system instead the container is vibrated with a sinusoidal
law for its top vertical position $A\sin(2\pi\nu t)$, with a constant
frequency $\nu=200$ Hz and amplitude $A \in [0.03,0.25]$ mm. We recall
that the maximum vertical acceleration (divided by gravity $g$)
$\Gamma$ varies in the range $20-40$ and the maximum vertical velocity
in $80-300$ mm/s. The effect of the shaking is the fluidization of the
granular medium, which - depending on $N$ and $\Gamma$ - stays in a
steady gas or liquid regime. The probe for diffusion is a blade
(dimensions $35\times 6\times 15$ mm, momentum of inertia $I = 353$ g
mm$^2$) mechanically isolated from the container, that can rotate
around a centered vertical axis and takes energy only from collisions
with the granular medium. The angular velocity $\omega(t)$ of the
blade and its absolute angle of rotation $\theta(t) = \int_0^t
\omega(t')dt'$ are measured in the experiment by an encoder with high
spatial and temporal resolution (see Supplemental Materials
in~\cite{scalliet}). The numerical simulations are performed with a
discrete-elements model implemented by LAMMPS
package~\cite{plimpton1995fast} with interactions obeying a nonlinear
visco-elastic Hertzian model that takes into account the relative
deformation of the grains.  In order to properly
reproduce the soft collision dynamics,  simulations run with a time step of $\sim 10^{-5}$ s. Our macroscopic phenomena occur on time-scales larger than $10$ s, therefore we integrate the dynamics for total times $t_{TOT}= 3.6 \cdot 10^3$ s, spanning more than eight decades of time-steps.
More details about the simulation, including the parameters that
control the dissipative and the elastic contributions of the
interaction, are given in the Supplemental Material (SM)~\cite{sm}.
There the reader can also find a table with the correspondence between
$N$ and best estimates of the packing fraction $\phi$. We stress that,
in view of the collective nature of the phenomena studied here, it is
reasonable to expect a certain robustness of the results with respect
to the microscopic details. Indeed, the main experimental phenomena
are qualitatively reproduced in large regions of the parameters'
space. The effects of changing the dissipation of interactions, the
mass or the diameter of particles are discussed in the SM~\cite{sm}. Here
we focus on the particular set of values exhibiting the optimal
quantitative agreement (within a tolerance of roughly $10\%$) with the
experiment.


\begin{figure}[h!]
  \parbox{0.48\columnwidth}{
  \includegraphics[width=0.48\columnwidth]{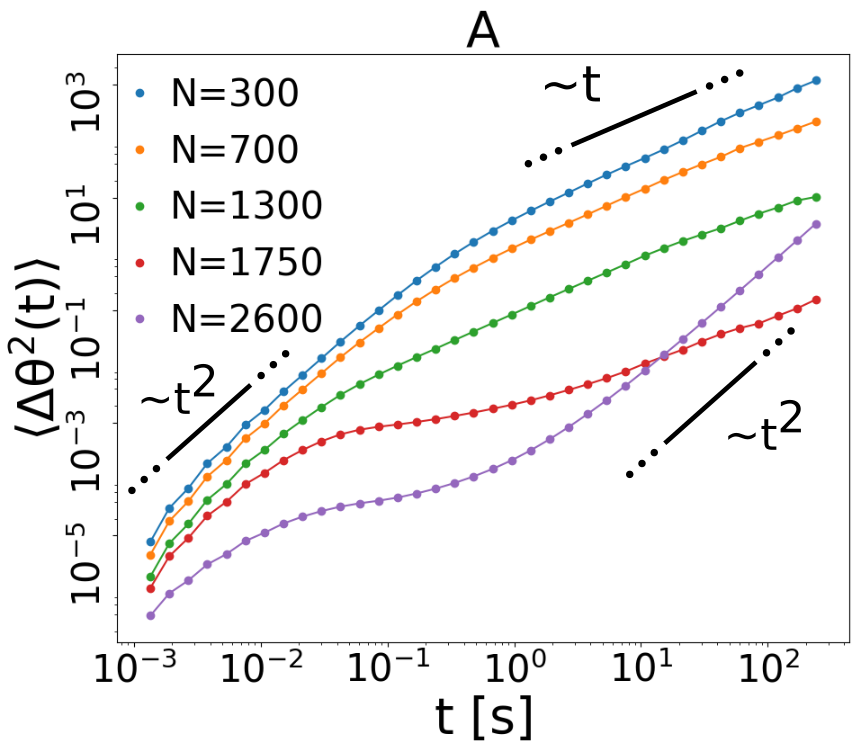}
  \includegraphics[width=0.48\columnwidth]{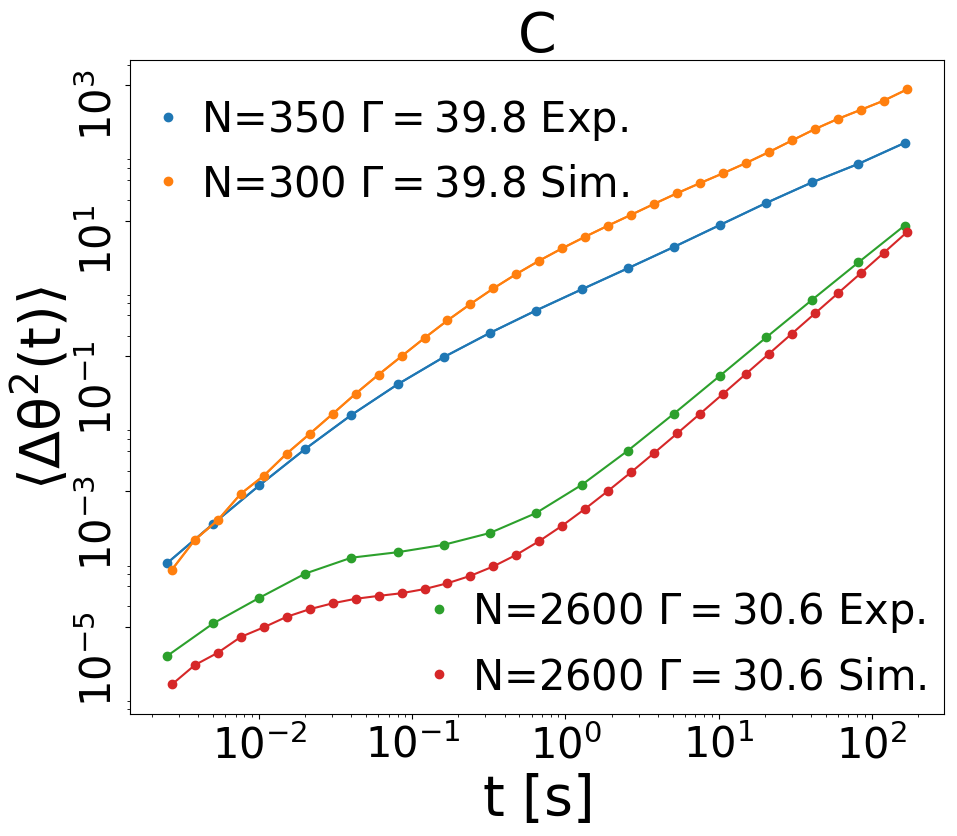}    }
  \parbox{0.48\columnwidth}{

    \includegraphics[width=0.48\columnwidth]{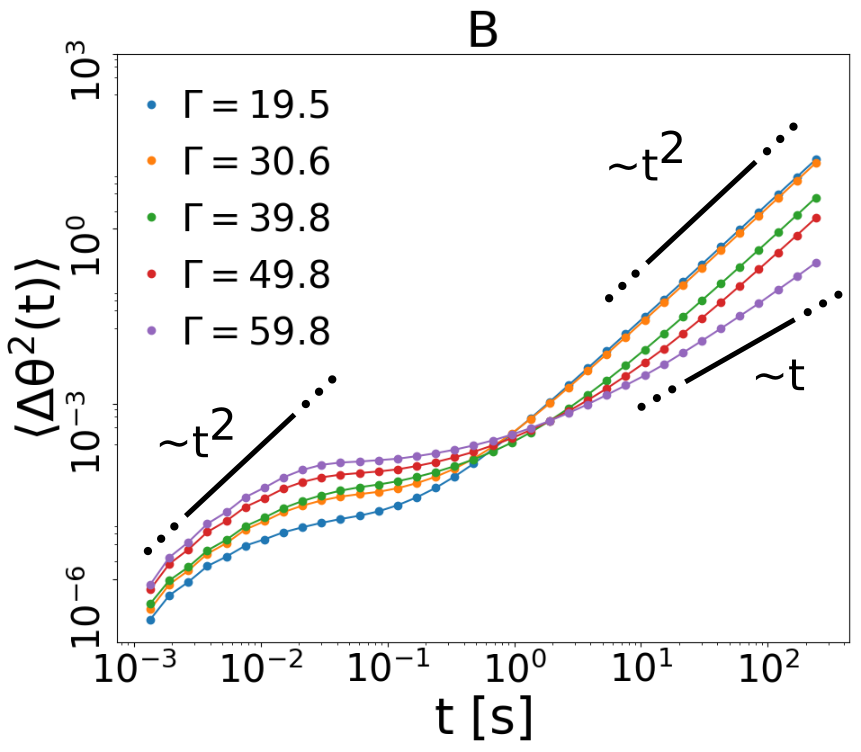}
    \includegraphics[width=0.48\columnwidth]{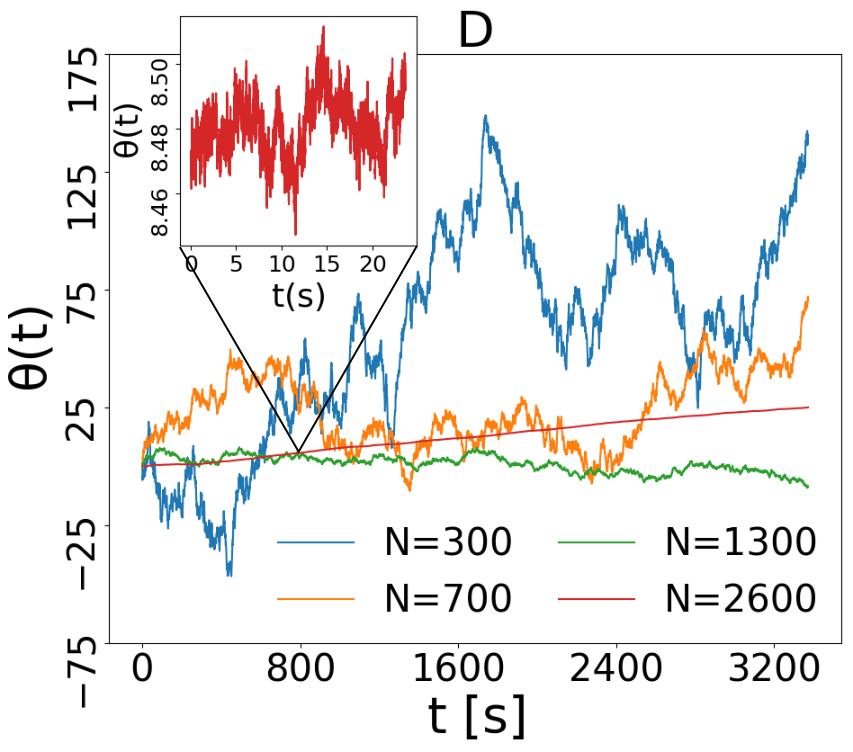}
  }
  \caption{Dynamics of the probe. A: msd in simulations with
    $\Gamma=39.8$ and different values of $N$. B: msd in simulations
    with $N=2600$ and different values of $\Gamma$. C: msd, comparison
    with experiments.  D: trajectories $\theta(t)$ in several
    simulations with $\Gamma=39.8$ and different values of $N$. \label{fig:compare}}
\end{figure}

{\em Numerical results: study of the probe.}  We first compare
numerical and experimental results for velocity fluctuations $\langle
\omega^2 \rangle-\langle \omega\rangle^2$ (proportional to probe's
kinetic temperature), not shown here, which fairly agree in the whole
range of values of $N$ and $\Gamma$. In Fig.~\ref{fig:compare} A and B
we show the results of the crucial test of our simulation, i.e. the
msd in many different situations. When $N$ is small and $\Gamma$ is
high (cyan, yellow and green curves in frame A) the granular medium
provides the blade with uncorrelated impacts and the large mass of the
probe is sufficient to predict a leading order Ornstein-Uhlenbeck
behavior~\cite{sarra10,GPT13}: this determines the typical ballistic
($msd \sim t^2$) to diffusive ($msd \sim t$) scenario. When $N$
increases the medium entraps the probe - similarly to the cage effect
in molecular liquids - resulting in a transient arrest of the msd at
intermediate times $t \sim t_{cage}$, followed by the usual ``escape''
with the behavior $msd \sim t$ when $t \gg t_{cage}$ (red curve in
frame A, purple curve in frame B). A peculiar granular feature is the
possibility to develop late-time superdiffusive behavior, when $N$ is
increased further (purple curve in frame A) or $\Gamma$ reduced
further (cyan, yellow, green and red curves in frame B). At the
smallest values of $\Gamma$ we observe the limit case $msd \sim t^2$
when $t \gg t_{cage}$. In Fig.~\ref{fig:compare}C we demonstrate the
good comparison, appreciable also at the quantitative level, between
the msd observed in the simulations and in the experiments, in two
very different cases.  Finally, in Fig.~\ref{fig:compare}D we show
some examples of the numerical time-series $\theta(t)$ (along $\sim 1$
hour) when $N$ is tuned from the dilute gas to the dense liquid
(constant $\Gamma$).  This plot reveals the
significant change in the probe's dynamics induced by the variation of
surrounding granular density. We remind that dissipative 
interactions naturally reduce velocity fluctuations
when $N$ is raised at constant $\Gamma$. Besides this, a non-trivial
memory effect clearly emerges, with long relaxation times evident in
the persistence of the direction of motion. 

\begin{figure}
  \includegraphics[width=0.94\columnwidth]{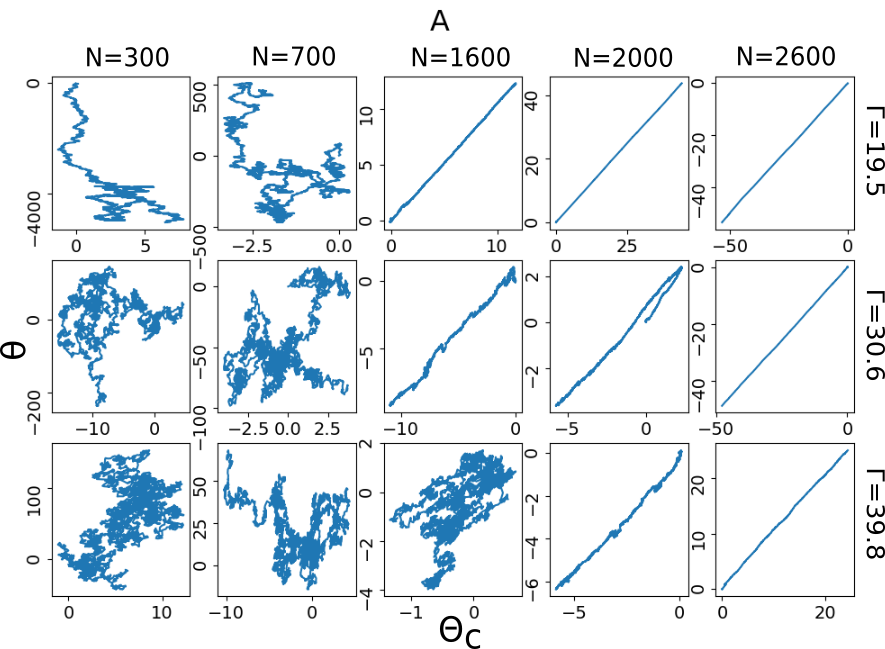}
  \includegraphics[width=0.48\columnwidth]{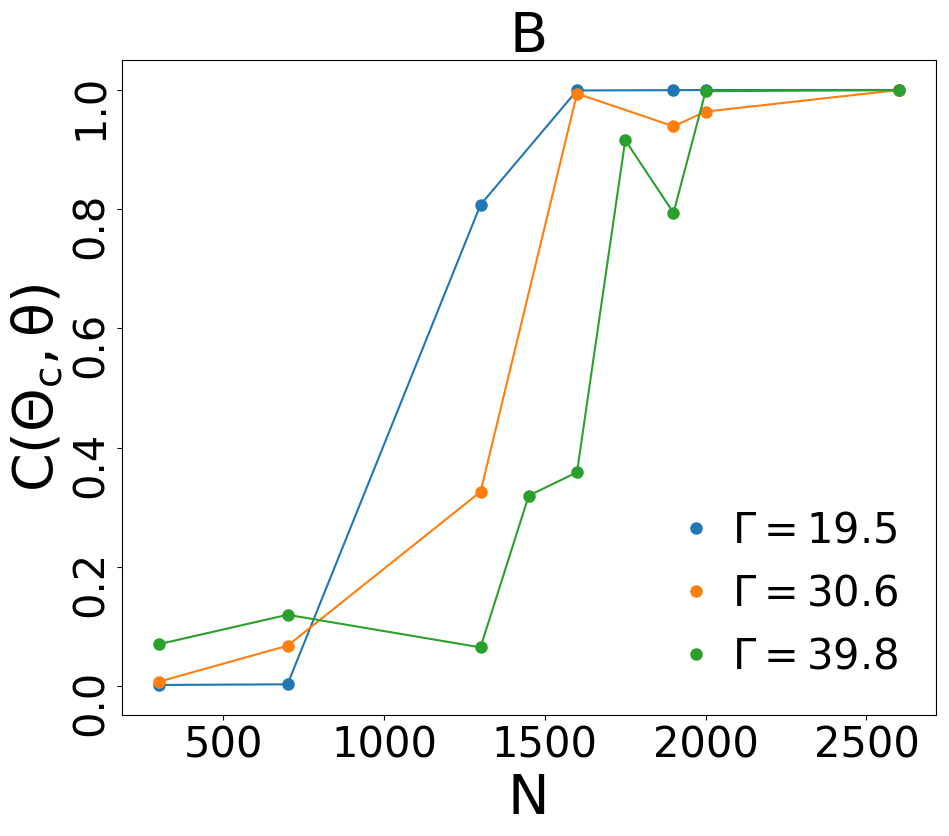}
  \includegraphics[width=0.48\columnwidth]{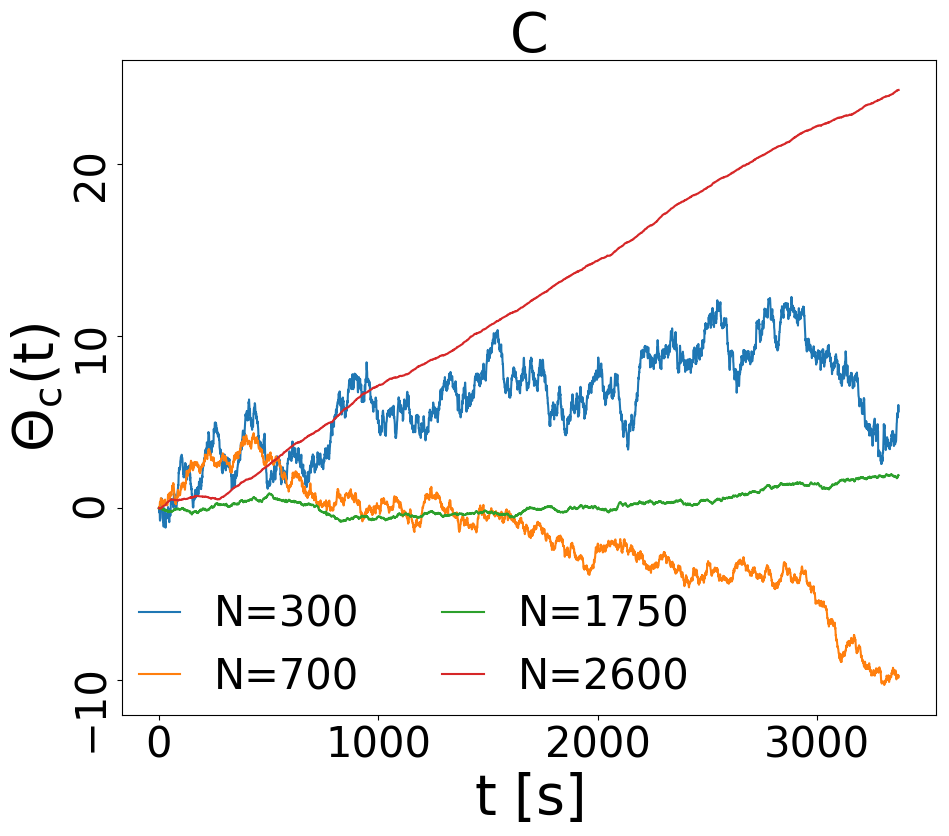}
  \caption{Comparison probe vs. collective rotation of the granular
    medium. A: parametric plot of the two rotations for several
    choices of $N$ and $\Gamma$. B: correlation between the
    probe and the collective rotation as a function of $N$ for
    different $\Gamma$. C: simulation without the probe,
    absolute collective angle $\Theta_c$ vs. $t$ for different $N$ at
    $\Gamma=39.8$. \label{fig:colpal}}
\end{figure}

{\em Numerical results: study of the medium.} Once the simulation has been successfully compared with the
experimental results, we can obtain new
information which was unreachable experimentally, due to the 3D nature
of the setup. Our first focus is on the most natural collective
variables which could be coupled to the angular velocity of the blade,
that is the average angular velocity of the granular medium (with respect to the central axis)
\begin{subequations}
  \begin{align}
\Omega_c(t)&=\frac{1}{N}\sum_{i=1}^N \dot{\theta}_i(t)\\
\theta_i(t)=\arctan\left(\frac{y_i(t)}{x_i(t)}\right) &\;\;\;\; \dot{\theta}_i(t)=\frac{({\mathbf r}_i(t) \times {\mathbf v}_i(t))_z}{r_i^2},
  \end{align}
\end{subequations}
and its time-integral which represents a collective absolute angle
$\Theta_c(t)=\int_0^t \Omega_c(t') dt'$.

A solid evidence that $\Theta_c(t)$ is meaningful with respect to the
anomalous diffusive properties of the probe can be found in
Fig.~\ref{fig:colpal}. Frame A reports the parametric plot
$\theta(t)$ (blade's angle) vs. $\Theta_c(t)$ (collective angle). It
is immediately clear that a sort of crossover line exists in the plane
of parameters $N,\Gamma$ (with, in fact, a weak dependence on
$\Gamma$ coordinate): for small values of $N$ and high values of
$\Gamma$ one has a phase where $\theta(t)$ and $\Theta_c(t)$ are
mostly uncorrelated. On the contrary, for large $N$ and small $\Gamma$
a strong correlation emerges between the two signals. In Fig.~\ref{fig:colpal}B we show a covariance
$C(\Theta_c,\theta)$ defined as
\begin{equation}
C(x,y)=1-\frac{\langle [x'(t)-y'(t)]^2\rangle}{\langle (x'(t))^2 \rangle+\langle (y'(t))^2 \rangle},
\end{equation}
with $x'(t)=x(t)-\langle x \rangle$, $y'(t)=y(t)-\langle y \rangle$
and the average runs over data sampled for the whole simulation time (typical simulation times are 3600 s with a temporal step $\Delta t=1.35\cdot 10^{-5}$ s).  Note that this
estimator of the covariance can also be rewritten as $C(x,y)=2\langle
x'(t)y'(t)\rangle/(\langle (x'(t))^2 \rangle+\langle (y'(t))^2
\rangle)$.  Frame B of Fig.~\ref{fig:colpal} confirms that it
can be interpreted as an order parameter distinguishing between those
two phases, and that the crossover occurs, with a fair sharpness, at a value $N_c \sim 1300 \div 1600$ only weakly
dependent upon $\Gamma$. 

The results of the previous analysis make clear that in the dense/cold
cases, when seen from the point of view of large time-scales ($\gtrsim
1 s$, roughly speaking), the dynamics of the blade is strongly
correlated to the collective rotation of the granular medium, which is
the real hallmark of the transition. This suggests us to focus on a
new series of simulations {\em without} the rotating blade, with the
aim of focusing upon the granular medium itself~\footnote{Our analysis
  revealed that the behavior of the granular medium is largely
  independent from the presence or absence of the blade, however a
  blade-free simulation seemed us {\em cleaner} with respect to
  assessing the minimal ingredients of the observed dynamics.}.  The
analysis of $\Theta_c(t)$ in these ``blade-free'' simulations, shown
in a few relevant cases in Fig.~\ref{fig:colpal}C, immediately
corroborates this intuition: the collective absolute angle of the
granular medium is erratic and Brownian-like for $N<N_c$ and much more
smooth at $N>N_c$. In particular the cases at $N>N_c$ appear
constituted by long periods where $\Theta_c$ travels in a constant
direction, interrupted by rare turns. The average time $t_{coll}$
between those turns seem to increase with $N$, up to a point (case at
the largest available $N=2600$, red curve) where it is longer than the
whole simulation. We also run longer simulations, not shown here,
confirming that sudden turns occur also at $N=2600$, but with
$t_{coll} \gg 10^3$ seconds. The interesting connection between
spatial rearrangements and changes of rotation speed or direction has
eluded our attempts and remains an open question for future
investigations.

\begin{figure}
  \includegraphics[width=0.49\columnwidth]{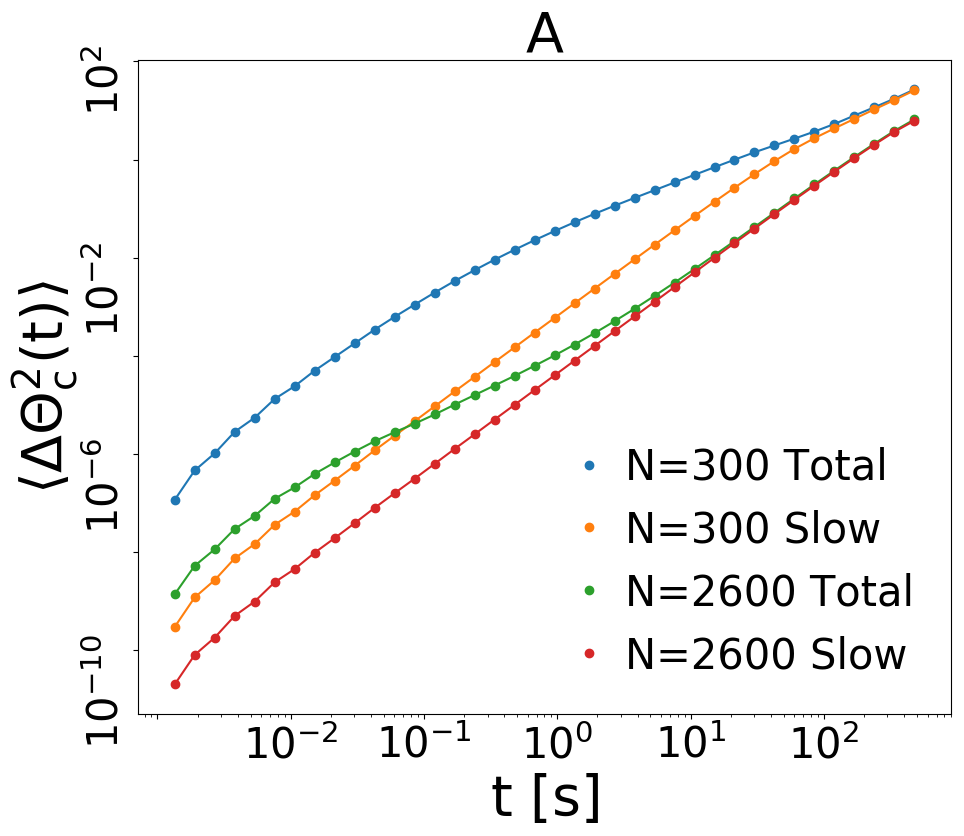}
      \includegraphics[width=0.49\columnwidth]{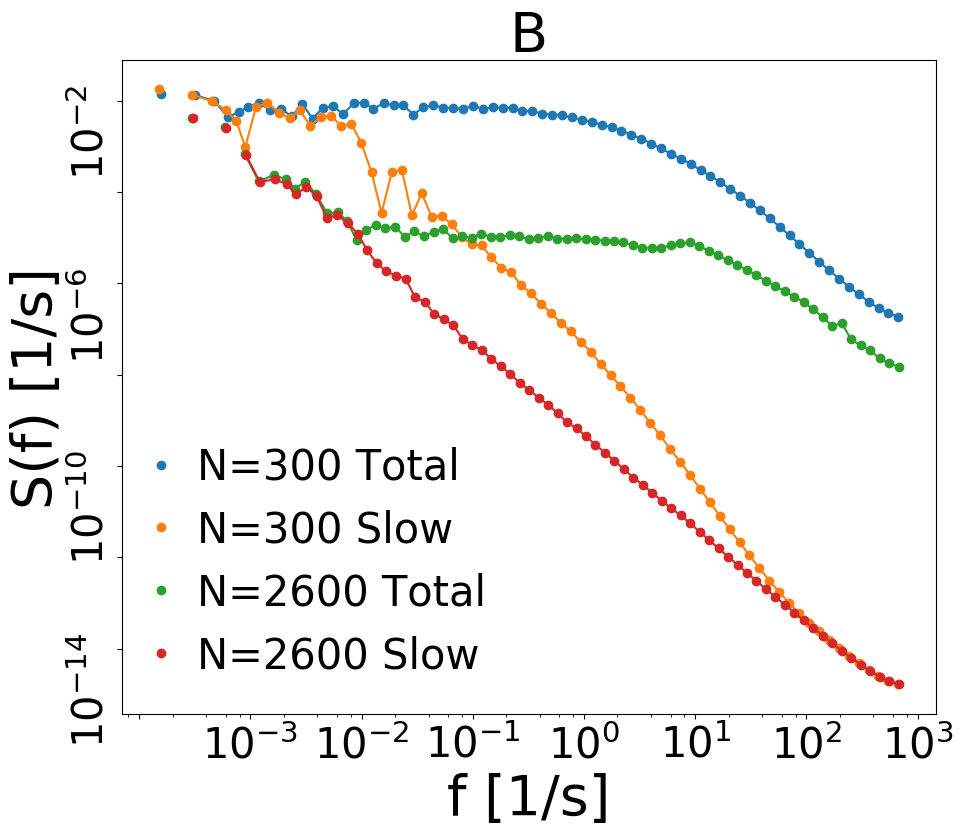}
  \caption{Collective variable $\Theta_c(t)$ at two different values
    of $N$ and $\Gamma=39.8$, compared with its ``slow component'', see text for
    definition. A) Msd; B) Power spectra of $\Omega_c$. \label{fig:spectrum}}
\end{figure}

As expected from the strong correlation between $\theta(t)$ and
$\Theta_c(t)$, for $N>N_c$, the large-time behavior of the msd of
$\Theta_c(t)$ is superdiffusive exactly as already seen for the probe,
see Fig.~\ref{fig:spectrum}A. We have also verified, see SM~\cite{sm},
that the distributions of angular displacements at times larger than
$t_{cage}$ are always close to Gaussian, ruling out large tails such
as those in~\cite{biroli2010}. Superdiffusive behavior is therefore
caused by the long persistent angular drifts discussed before.
The connection is the following: in the dense cases $t_{coll}$ becomes
huge; therefore what we call ``large times'', with respect to
microscopic and intermediate ($t_{cage}$) timescales, are actually
``small times'' with respect to $t_{coll}$. Only following the signal
for times much longer than $t_{coll}$ one would recover the asymptotic
(or ``final'') diffusive behavior $msd \sim t$. In order to focus on
the dynamics at time-scales $\sim t_{coll}$, we get rid of the smaller
time-scale $t_{cage}$ by defining the slow collective velocity
  $\Omega_s(t)=\frac{1}{\tau}\int_t^{t+\tau} \Omega_c(t')dt'$,
with $\tau$ larger than $t_{cage}$ (for instance $\tau=1.35$
s). Defining, analogously, the slow component of $\Theta_c(t)$,
i.e. $\Theta_s(t)=\int_0^t \Omega_s(t') dt'$ gives also the
possibility of looking at the collective msd when the fast position
oscillations are filtered out, see again Fig.~\ref{fig:spectrum}A. The
slow component, surprisingly, has always a persistent ballistic-like
motion, even in the most dilute cases. However, in the dilute cases
this ballistic motion is a very weak signal and does not appear in the
msd of the total variable. The same observation can be seen for the
power spectrum of the angular velocity $\Omega_c$ which is defined as $S(f)=\frac{1}{2\pi t_{TOT}}\lvert\int_0^{t_{TOT}}
\Omega_c(t) e^{i (2\pi f) t}dt \rvert^2 $~\cite{scalliet}, shown in
Fig.~\ref{fig:spectrum}B. In both dilute and dense cases the slow
component (yellow and red curves) has a Lorentzian-shaped spectrum,
with a flat part at small frequencies $f$ and a decaying part, roughly
$\sim f^{-2}$ at larger frequencies. Only in the dense case $N=2600$
such a low-frequency ``flat'' part intervenes at very small
frequencies, so that part of $\sim f^{-2}$ emerges in the total
signal. We verified that changing $\tau$ around a value slightly
larger than $t_{cage}$ does not affect this scenario.

\begin{figure}[t]
  \includegraphics[width=0.49\columnwidth]{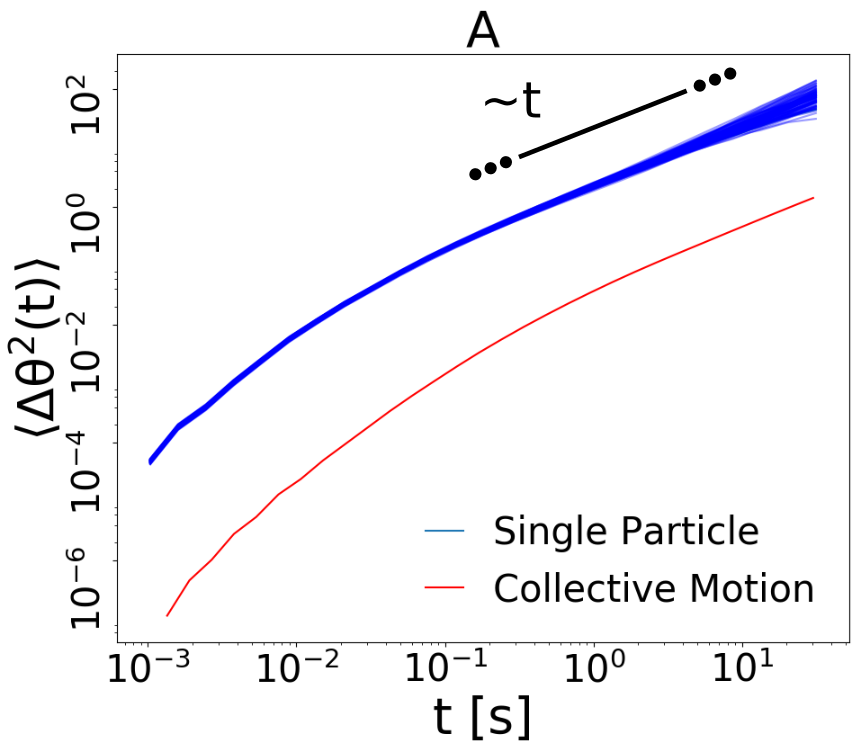}
  \includegraphics[width=0.49\columnwidth]{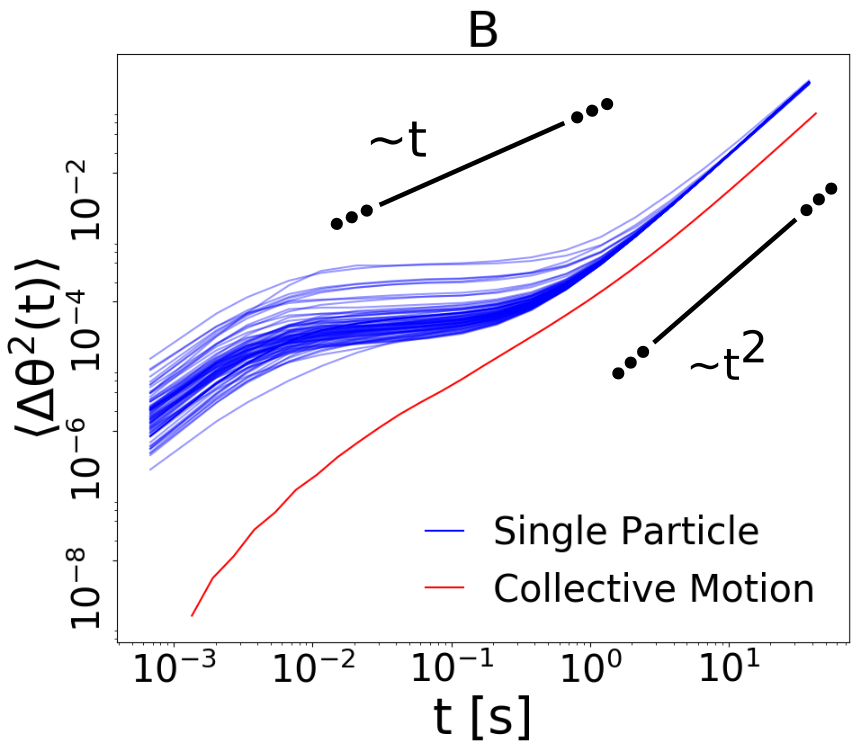}
  \caption{Mean squared displacement of $\sim 100$ particles spread randomly in the system, in a dilute case and a dense case, with same parameters as those in Fig.~\ref{fig:spectrum}.  \label{fig:etero}}
\end{figure}

A last question which we address concerns fluctuations, which may be
crucial to define a proper correlation length. In glassy systems a
well established route is to study dynamical heterogeneities,
i.e. wide variations in the local mobility of the medium between
different
samples~\cite{ediger2000spatially,berthier2011dynamical,dauchot2005dynamical,durian07,candelier2009building}. A
first encouraging information comes by analyzing the difference in the msd
of $100$ particles randomly picked, uniformly in space, in the whole
system. Fig.~\ref{fig:etero} indicates that the
dense case, with respect to the dilute one, exhibits a much wider
volatility of msd.  Such heterogeneity is large at intermediate times
$t\sim t_{cage}$ and rapidly decreases at late times, i.e. when
superdiffusion sets in. This confirms that superdiffusion is related
to a strong coordination in the trajectories of particles.

In conclusion, we demonstrate that probe's
anomalous diffusion originates in a collective mode of the host
granular medium, even if it is constituted of many non-polar
particles. The slow component of the grains' movement along a
direction which is free of obstacles, i.e. rotation around the central
axis, is more and more coordinated and persistent as density increases
and temperature decreases. The observed strong and enduring
correlations are not particularly sensitive to changes in the
dissipation and in the size of the system and, noticeably, to the
degree of configurational (e.g. crystalline) order, see
SM~\cite{sm}. The robustness of our results suggests that they could
apply to other systems, for instance granular materials under
different conditions of fluidizations, as well as disordered glassy
systems and fluids of active particles. The dynamical memory displayed
by our system could be a more easy to observe proxy for very slow
configurational relaxation in hard sphere glasses~\cite{Gradenigo10}.
Moreover, persistent alignment of velocities of inertial active
particles is observed in animal collective behavior, such as
in flocks~\cite{cavagna2018physics}. While its theoretical connection with granular materials
is established~\cite{manacorda17,manacordabook}, our work supports the experimental grounds for
the development of a common physical understanding of such diverse systems.



\bibliography{biblio}

\newpage

\onecolumngrid

{\bf SM: SUPPLEMENTAL MATERIAL}

\section{Interaction model for the grain contacts}

Our system is simulated through the LAMMPS package~\cite{plimpton1995fast}. The Hertz-Mindlin model is used
\cite{Zhang2005,Silbert2001,Brillantov1996} to solve the
dynamics during the particle-particle and particle-container contact. This visco-elastic
model takes in account both the elastic and the dissipative response
to the mutual compression between the grains. In addition, it includes
in the dynamics not only the relative translational motion but also the
rotational one. The modeled forces are described by Eq. \ref{ForzaHM} below. In
Fig. \ref{MovieUrto} we can understand the physical meaning of the
variables in play.

\begin{equation}\label{ForzaHM}
\begin{cases}
    \vec{F}^{N}_{ij}= \sqrt{R_{ij}^{\text{eff}}}\sqrt{\xi_{ij}(t)}\left[(k_{n}\xi_{ij}(t)-m^{\text{eff}}\gamma_{n}\dot{\xi}_{ij}(t))\cdot
\vec{n}(t)\right] \\
    \vec{F}^{T}_{ij}=\vec{F}^{\text{hist}}_{ij} \; \; \; \text{if} \; \; \; |\vec{F}^{\text{hist}}_{ij} | \le |\mu\vec{F}^{N}_{ij}|     \\
\vec{F}^{T}_{ij}=-\dfrac{|\mu\vec{F}^{N}_{ij}|}{|\vec{g}^{T}_{ij}(t)|}\cdot \vec{g}^{T}_{ij}(t) \; \; \; \text{otherwise} \\
\vec{F}^{\text{hist}}_{ij}=-\sqrt{R_{ij}^{\text{eff}}}\left[k_t\bigintssss_{\text{s(t)}}\sqrt{\xi_{ij}(t')}\vec{ds}(t')+m_{\text{eff}}\gamma_{t}\sqrt{\xi_{ij}(t)}\vec{g}^{T}_{ij}(t)\right]
   
\end{cases}
\end{equation}

\begin{figure}
\center \makebox[\textwidth][c]{
  \includegraphics[width=0.8\textwidth]{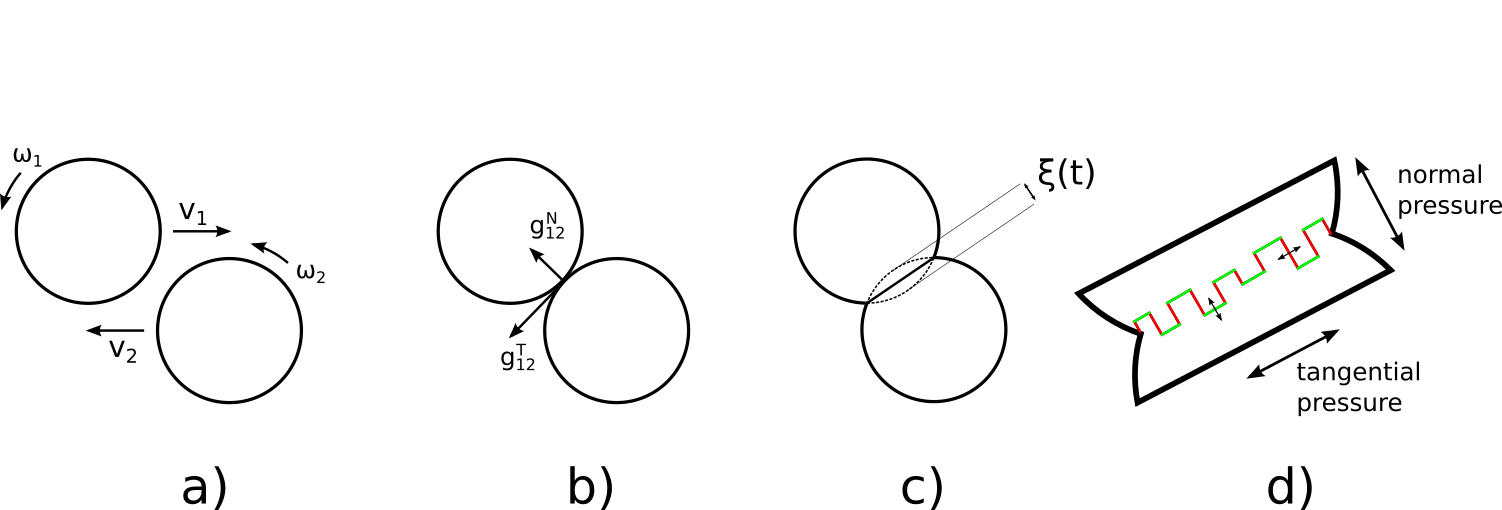}}\caption{a)
  Two grains are going to collide with given linear and angular
  velocities. b) The contact between the grains starts with a normal
  and tangential relative velocity $\vec{g}^{N}_{12}$ and
  $\vec{g}^{T}_{12}$. c) An example of surface deformation and
  definition of the instantaneous normal compression $\xi(t)$. d)
  Microscopic interpretation of the model: superficial asperities define a
  normal and a tangential surfaces of contact that enable respectively
  the transmission of impulse and angular momentum~\cite{Brillantov1996}.}
\label{MovieUrto}
\end{figure}

The equations and the figure are written for two particles with radius
$R_i$, $R_j$, positions $\vec{r}_i$, $\vec{r}_j$, translational
velocities $\vec{v}_i$, $\vec{v}_j$ and rotational velocities
$\vec{\omega}_i$, $\vec{\omega}_j$.  The relative velocity is defined
as $\vec{g}_{ij}= (\dot{\vec{r}}_{i}-\vec{\omega}_{i}\times
R_{i}\vec{n})-(\dot{\vec{r}}_{j}+\vec{\omega}_{j}\times R_{j}\vec{n})$
where
$\vec{n}=\left(\vec{r}_{i}-\vec{r}_{j}\right)/\left|\vec{r}_{i}-\vec{r}_{j}\right|$ defines the normal direction;
we call $\vec{g}_{ij}^{N}$ and $\vec{g}_{ij}^{T}$ the two projections,
respectively normal and tangential, to the surface of contact.  The
instantaneous normal compression is represented by
$\xi_{ij}(t)=R_{i}+R_{j}-|\vec{r}_i-\vec{r}_{j}|$ and its derivative
is $\dot{\xi}_{ij}(t)=|\vec{g}^{N}_{ij}|$.  During the contact, the
particles are subjected to a normal force $\vec{F}^{N}_{ij}$ and a
tangential one $\vec{F}^{T}_{ij}$; both these components have an
elastic and a dissipative contribution individuated by
the couples of parameters $k_{n}$, $k_{t}$ and $\gamma_{n}$, $\gamma_{t}$, respectively.  In the
normal force $\vec{F}^{N}_{ij}$ we can see an elastic term that
descends from the Hertzian theory of contact mechanics
\cite{PopovBook} characterized by a non-linear dependence on the
displacement. We recall that in the framework of the same theory it is also possible to
derive the dissipative term \cite{Brillantov1996}. The less intuitive
term of the model is surely the elastic-tangential one because it
depends upon the memory force $\vec{F}^{\text{hist}}_{ij}$ that takes into
account the past history of the tangential displacement $s(t)$
\cite{Zhang2005}. We finally remark that in this model the tangential
force is limited by Coulomb friction, characterized by the
coefficient $\mu$ (Eq.\ref{ForzaHM}).

\section{Details of the simulation}
Now we briefly discuss how we set the interaction parameters and the simulation time step in relation to the properties of the materials in play. The elastic coefficients can be directly derived as 

\begin{equation} \label{CostEl}
k_{n}=\dfrac{2Y}{3(1-\nu^2)}, \; \; \; k_t=\frac{4Y}{(2-\nu)(1+\nu)},
\end{equation}
where $Y$ is the Young modulus and $\nu$ in the Poisson ratio. This holds for contacts between same material but it is possible to generalize to the case of two species \cite{Zhang2005,PopovBook,DiMaio2004}. Direct formulas for $\gamma_n$ and $\gamma_t$ are lacking but it is a common strategy to choose them verifying a posteriori the good agreement with the experimental data \cite{PoeschelBook}. 
The simulation time step has been chosen as a fraction of the Rayleigh time $dt=0.2t_{\text{ray}}$ namely the time that a superficial acoustic wave takes to cross a single grain. It is related to the characteristics of the material:

\begin{equation}
t_{\text{ray}}=\dfrac{\pi R_{\text{min}} \sqrt{2\left(1+ \nu\right)\rho/Y}}{0.163 \nu +0.8766 },
\end{equation} 
where $\rho$ is the grain density and $R_{\text{min}}$ the radius of
the littlest grain in the system \cite{Rackl2017}. This is a good
choice to properly resolve the dynamics because $t_{\text{ray}}$ is
usually much smaller than the typical collision times.  We finally
report the numeric values of the parameters in the
Tab. \ref{tab:TabParam} where superscript $a$ is referred to steel and
$b$ to Plexiglas. It is worth underlining that we choose a Young
modulus for the steel that is two order of magnitude smaller than the
real one ($\sim 10^{9}$ Pa). The reason is the need to have a bigger $dt$
in order to decrease the simulation time.
This is a general strategy in DEM simulations when experimental data are available and hence it is possible to verify that the softening of the grains doesn't affect the underlying phenomenology.

\begin{table}[h]

\centering
\begin{tabular}{|c|c|} 
\hline 
$Y^a$ & 210 Mpa \\ 
\hline 
$\nu^a$ & 0.293 \\ 
\hline 
$\mu^{aa}$ & 0.5\\ 
\hline 
$k_{n}^{aa}$ & 153$\cdot 10^{6} \text{pa}$ \\ 
\hline 
$k_{t}^{aa}$ & 355$\cdot 10^{6} \text{pa}$ \\
\hline
$\gamma_{n}^{aa}$ & 3$\cdot 10^{4} (\text{sm})^{-1}$ \\
\hline
$\gamma_{t}^{aa}$ & 1$\cdot 10^{4} (\text{sm})^{-1}$ \\
\hline
$Y^{b}$ & 33 Mpa \\
\hline
$\nu^{b}$ & 0.370 \\
\hline

\end{tabular} \quad
\begin{tabular}{|c|c|}
\hline
$Y^{ab}$ & 33 Mpa \\
\hline
$\mu^{ab}$ & 0.5 \\
\hline 
$k_{n}^{ab}$ & 611$\cdot 10^{5} \text{pa}$ \\ 
\hline 
$k_{t}^{ab}$ & 132$\cdot 10^{6} \text{pa}$ \\
\hline
$\gamma_{n}^{ab}$ & 3$\cdot 10^{7} (\text{sm})^{-1}$ \\
\hline
$\gamma_{t}^{ab}$ & 1$\cdot 10^{5} (\text{sm})^{-1}$ \\
\hline
$t_{\text{ray}}$ & 6.75$\cdot 10^{-5} \text{s}$ \\
\hline
$dt$ & 1.35$\cdot 10^{-5} \text{s}$ \\
\hline
\end{tabular}

\caption{Numerical values for the material properties and the coefficients of the visco-elastic interaction. We also report the timestep $dt$ used for the simulations.}
\label{tab:TabParam}

\end{table}

\section{Effective Packing Fraction Of the system}

Here we report a table (Tab. \ref{tab:packtab}) with the average 
packing fraction $\phi$ of the system corresponding to the most
representative values of $\Gamma$ and $N$.

\begin{table}[h]
\centering
\begin{tabular}{|c|c|c|c|c|c|c|} 
\hline 
\backslashbox{$\Gamma$}{N} & 300 & 700 & 1300 & 1600 & 2000 & 2600  \\ 
\hline 
19.5 & 13.8 & 24.6 & 39.3 & 45.0 & 51.6 & 56.3\\ 
\hline 
30.6 & 8.4 & 17.7 & 30.8 & 37.3&44.8&52.9\\ 
\hline 
39.8 & 6.4 & 14.5 & 26.7 & 32.8 & 41.1 & 48.8\\
\hline
\end{tabular}
\caption{Table of packing fraction (\%) for diffrent couple of the parameters $\Gamma$ and $N$}
\label{tab:packtab}
\end{table}



\section{Distribution of displacements}

\begin{figure}[htbp]
  \includegraphics[width=0.4\textwidth]{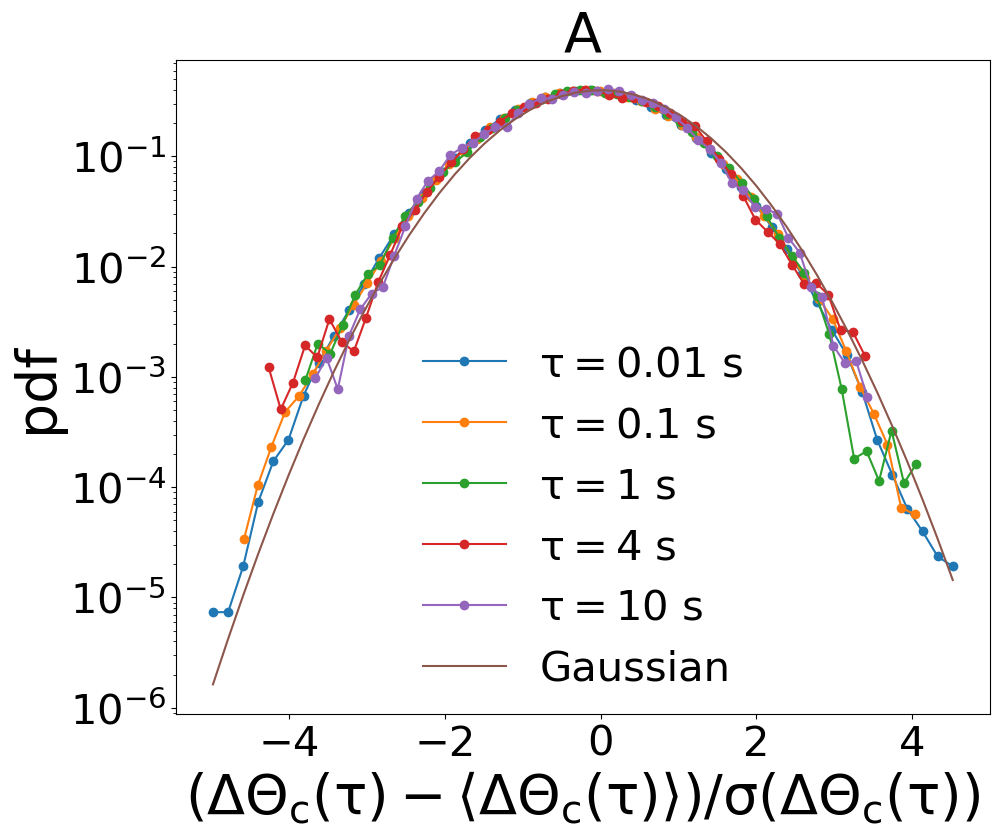}
  \includegraphics[width=0.4\textwidth]{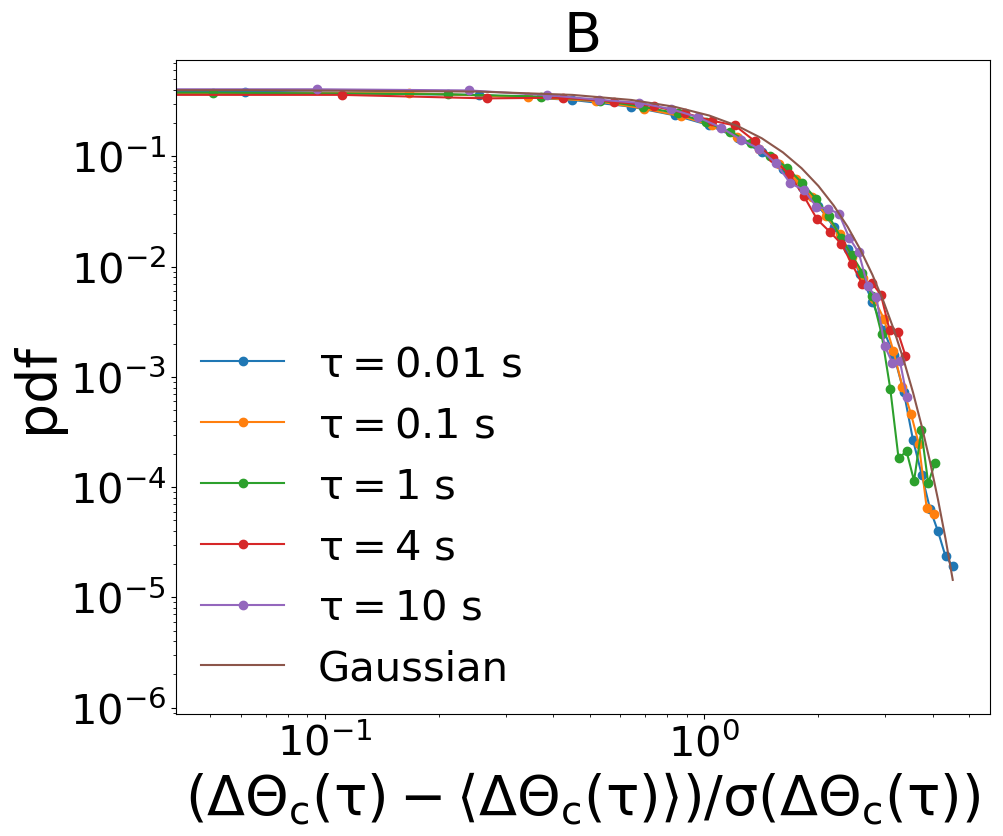} \\
  \includegraphics[width=0.4\textwidth]{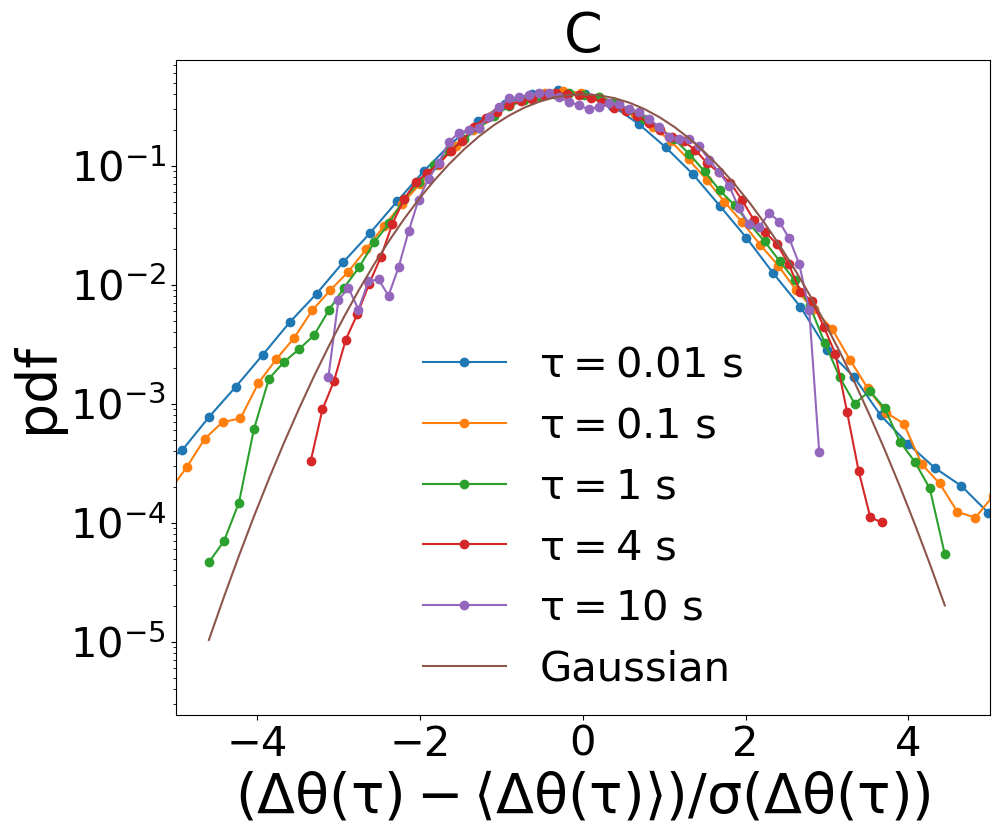}
  \includegraphics[width=0.4\textwidth]{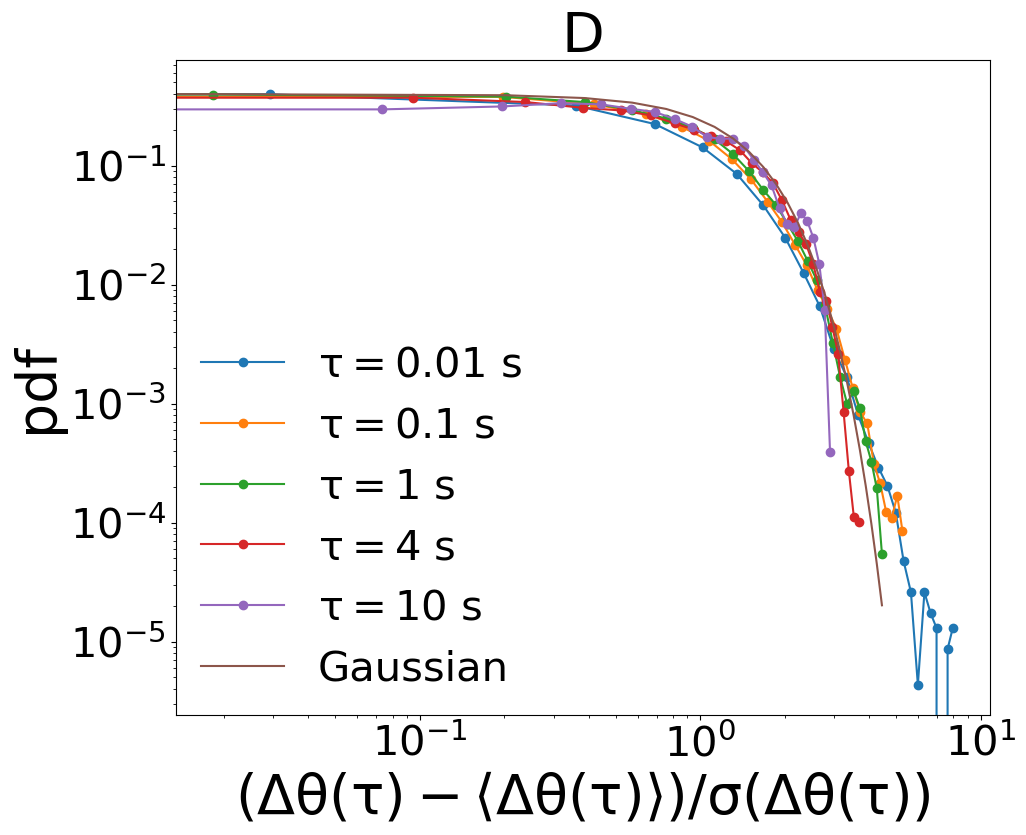}
\caption{Distributions of displacements for the  collective position  $\Theta_c(t)$ (upper row) and the single particle position  $\theta_i(t)$ (lower row) for delay times $\tau$. Distributions are collapsed by removing the average and rescaling by the standard deviation and shown in both semi-log and log-log scales. In all plots $\Gamma=39.8$ and $N=2600$. }
\label{pdfdispl}
\end{figure}

The superdiffusion observed for the single particle angular
displacement $\theta_i(t)$ as well as for the collective angular
variable $\Theta_c(t)$, see Figures 4A and 5B of the Letter, is
related to long persistence of almost ballistic flights in the system,
as documented by the trajectories of the variables (Figs. 1B, 3A and
3C). In order to strengthen our argument, we have measured also the
distributions of the displacements: this is useful to rule out an
alternative origin of the effect, that is the presence of large
(L\'evy) tails, such as those observed
in~\cite{biroli2010}. Distributions of the collective displacement,
shown in Fig.~\ref{pdfdispl}A and B, display a Gaussian shape at all delays
$\tau$. Distributions of the single particle displacement - see
Fig.~\ref{pdfdispl}C and D - on the contrary, have large tails (but
certainly with a finite variance) at small times $\tau< t_{cage}$ and
become Gaussian from times larger than $t_{cage}$.

\section{Effects of the main setup's parameters}

In this Section we explore the consequences of changing some
parameters of the numerical model, in particular: 1) we reduce the
particles' diameters (at constant density) in order to increase the
effective size of the system; 2) we add some polidispersity in the
system to rule out possible effects of crystallization; 3) we change
the dissipation in the interaction among the grains; 4) we change the
mass of the grains.

\subsection{System's size}

\begin{figure}[htbp]
  \includegraphics[width=0.4\textwidth]{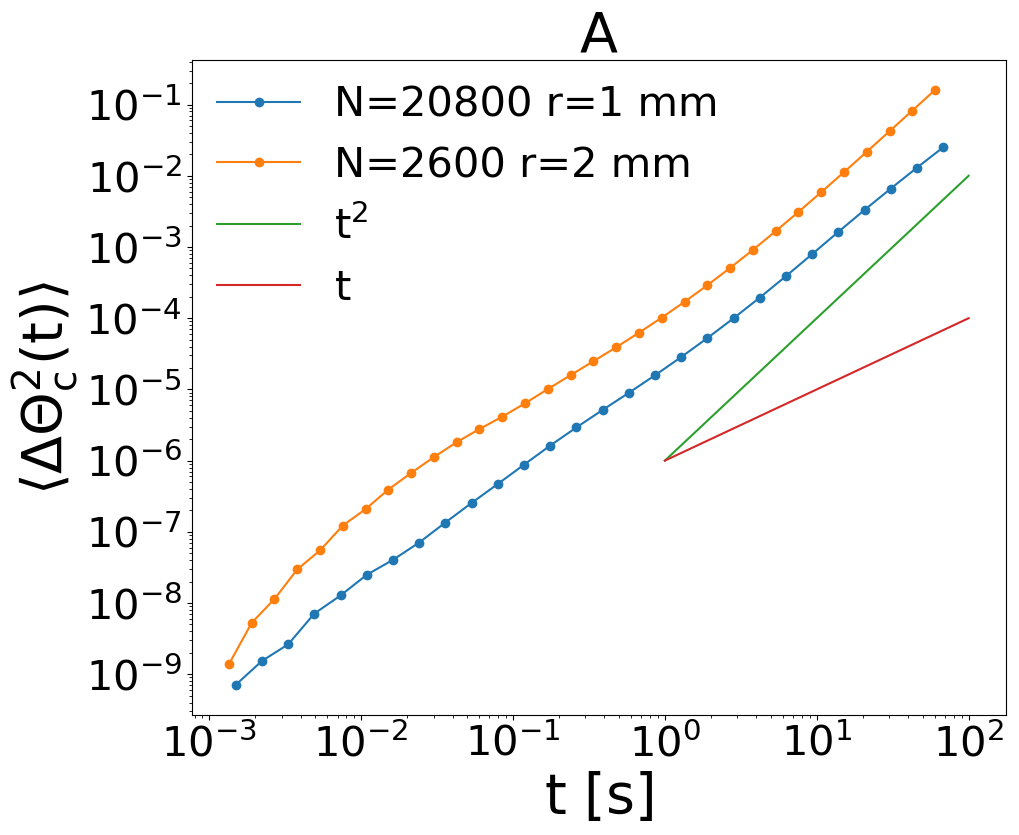}
  \includegraphics[width=0.4\textwidth]{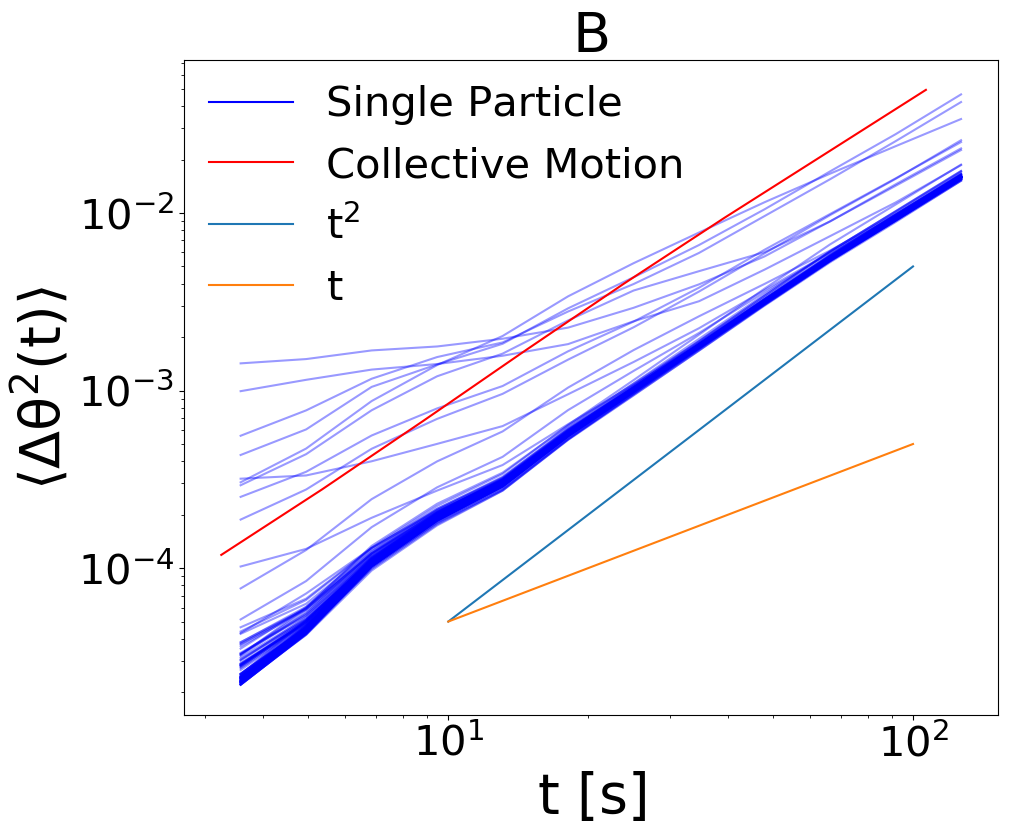}
\caption{Effect of reducing the diameter of the particles by a factor
  $2$ and increasing $N$ by a factor $2^3$, keeping constant
  $\Gamma=39.8$. The collective (frame A) and single-particle (frame
  B) mean-squared displacements of the angular coordinate are
  shown. The system with larger $N$ displays the same superdiffusive
  behavior $\langle \Delta \Theta_c^2 \rangle = \textrm{const} t^2$ at
  large times. }
\label{big}
\end{figure}

In Fig.~\ref{big} we show the mean squared angular displacement of the
collective variable (\ref{big}A) and of the single particles
(\ref{big}B) when the diameter of the particles is halved and the
number $N$ is increased by a factor $2^3$, so that the total average
packing fraction remains unaltered. Both collective and single
particle msd reveal superdiffusion at large times, with a power (close
to $\sim 2$) very similar to the smaller system. 

\subsection{Polidispersity and crystallization}

The long dynamical memory observed in both our real and numerical
experiments can be suspected of being related to a strong configurational order in the
system. In this subsection we rule out such a hypothesis by means of
two key observations. First we check that in our monodisperse
simulations the degree of order is below the crystallisation threshold
usually considered in the literature. Second, we run new simulations
with a polidisperse system, fairly reproducing the same superdiffusion
behavior observed in the monodisperse system.

\begin{figure}[htbp]
  \includegraphics[width=0.4\textwidth]{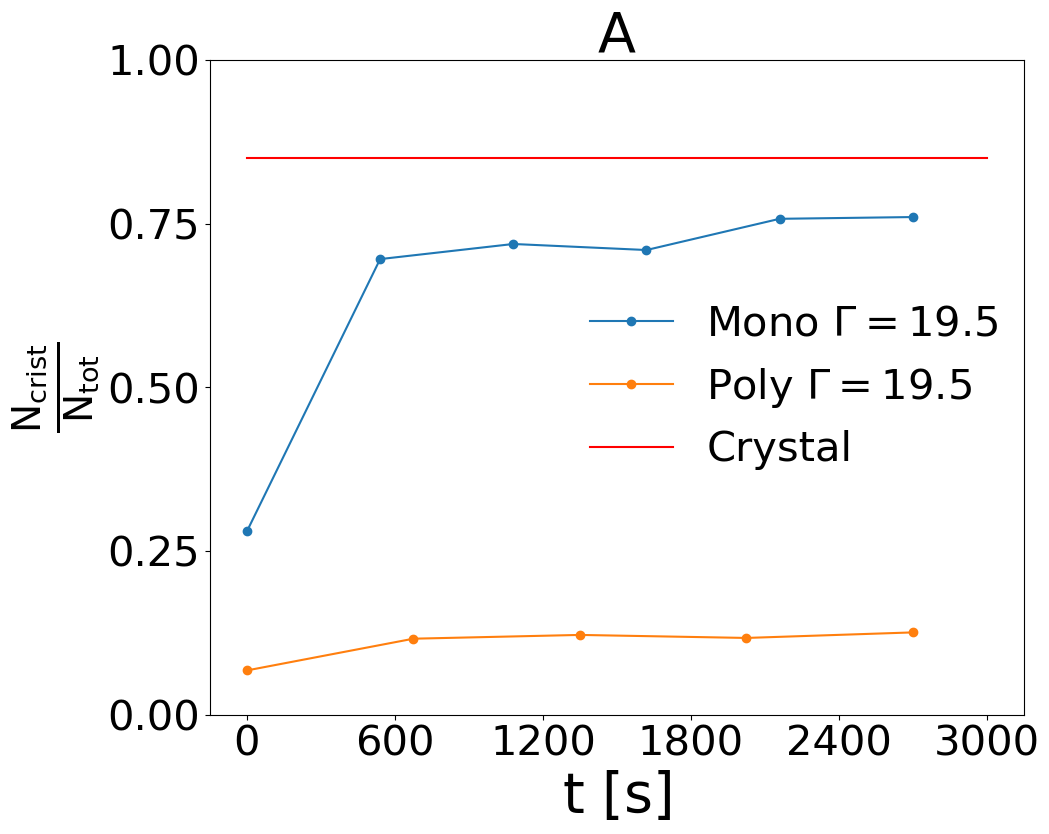}
  \includegraphics[width=0.4\textwidth]{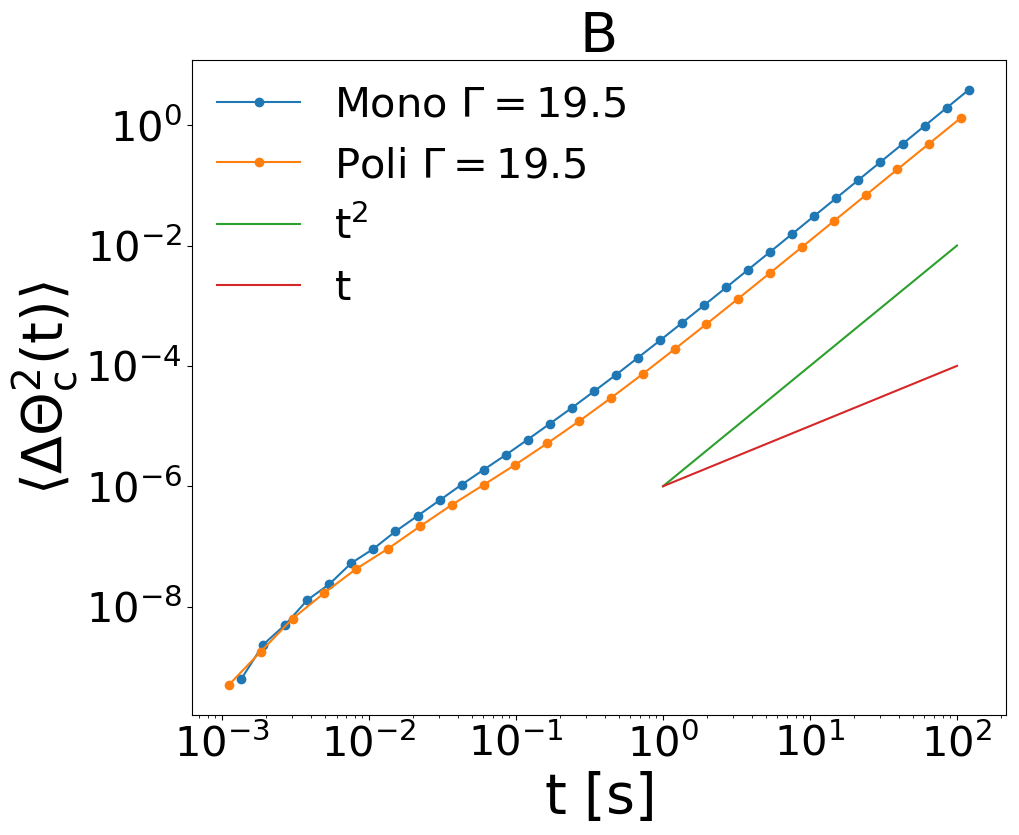}
\caption{Left: evolution with time of the percentage of crystallized
  particles (see text for the definition). Right: mean-squared angular displacement
  (collective variable) for a monodisperse and a polidisperse case
  with $\Gamma=19.5$. The polidisperse case is slightly colder than
  the monodisperse one, with an average kinetic energy per particle
  smaller of $\sim 20\%$.  }
\label{msdpoli}
\end{figure}

The degree of order in our system is measured by means of a
crystallographic order parameter defined in terms of spherical
harmonics, which is a usual and reliable observable to distinguish
between crystallized and disordered configurations of a many-particle
system, see for
instance~\cite{lechner2008,russo2012,zaccarelli2009,pusey2009}. For each particle $i$ a complex vector is defined as
\begin{equation}
q_{6m}^i = \frac{1}{N_b^i}\sum_{j=1}^{N_b^i} Y_{6m}(\hat{r}_{ij}),
\end{equation}
where $Y_{6m}$ are the 6th order spherical harmonics with $m$ an
integer that runs from $-6$ to $6$, $N_b^i$ is the number of neighbors
of particle $i$ (i.e. particles whose distance is smaller than $1.4$
radii) and $\hat{r}_{ij}$ is the unit vector joining particles $i$ and
$j$, defining a unique point on the sphere where the spherical
harmonic is defined. The scalar product ${\bf q}_{6}^i \cdot {\bf
  q}_{6}^j/(|{\bf q}_{6}^i||{\bf q}_{6}^j|)$ is computed and the
couple $ij$ is considered ``connected'' if this product exceeds
$0.7$. If the particle $i$ is connected to $6$ or more neighbors, then
it is considered crystallised. The fraction of crystallised particles
is shown in Fig.~\ref{msdpoli}A as a function of time, in order to
confirm that we are measuring a steady value. In the considered
monodisperse case (which is the densest in our series of
simulations) we measure a value smaller than or close to $0.75$, while
in the literature a configuration is considered a crystal when such a
fraction is larger than $0.85$.

Then we have run simulations with a significant degree of
polidispersity. This is obtained by placing in the container the
same number of particles but with random diameters, extracted according to a 9-component discrete Gaussian
distribution with average $4$ mm
and standard deviation equal to $10 \%$ of the average. The measure of
the crystallographic order, shown in Fig.~\ref{msdpoli}A demonstrates
that such a protocol guarantees a significantly smaller percentage of
crystallisation, $\sim 0.12$. Nevertheless, the behavior of the system, with
superdiffusive mean squared displacement, is fairly recovered, as
shown in Fig.~\ref{msdpoli}B. 

\subsection{Inelasticity and mass of the particles}

\begin{figure}[htbp]
  \includegraphics[width=0.4\textwidth]{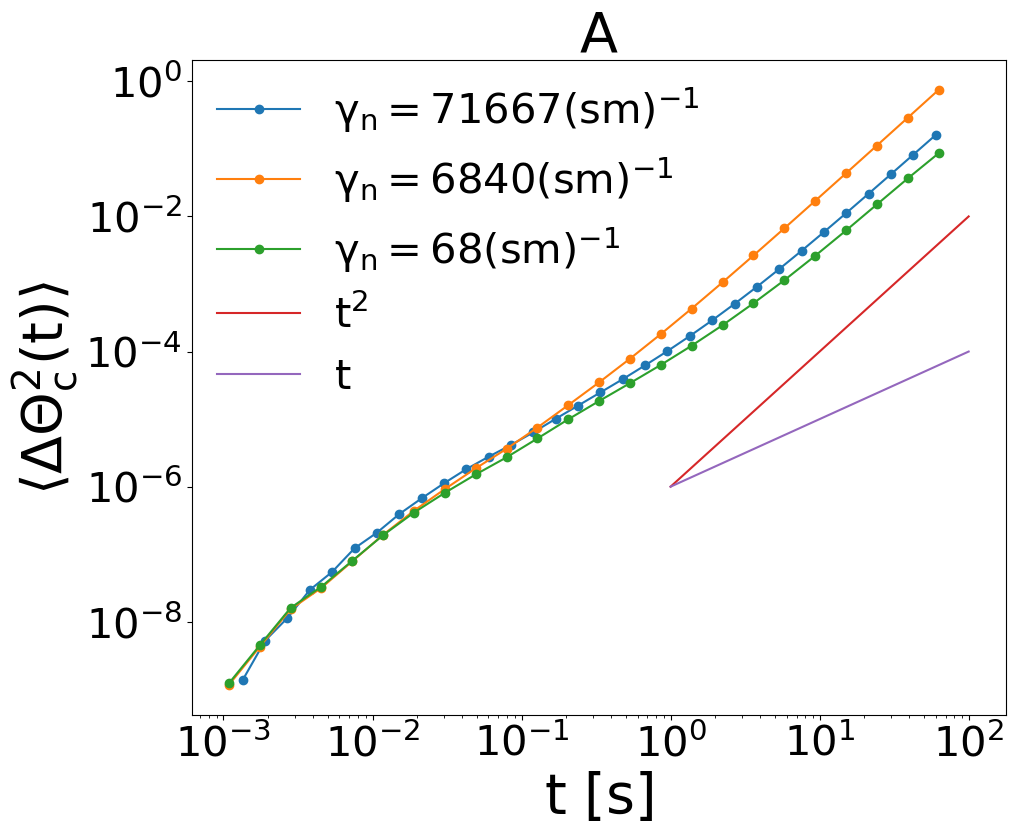}
  \includegraphics[width=0.4\textwidth]{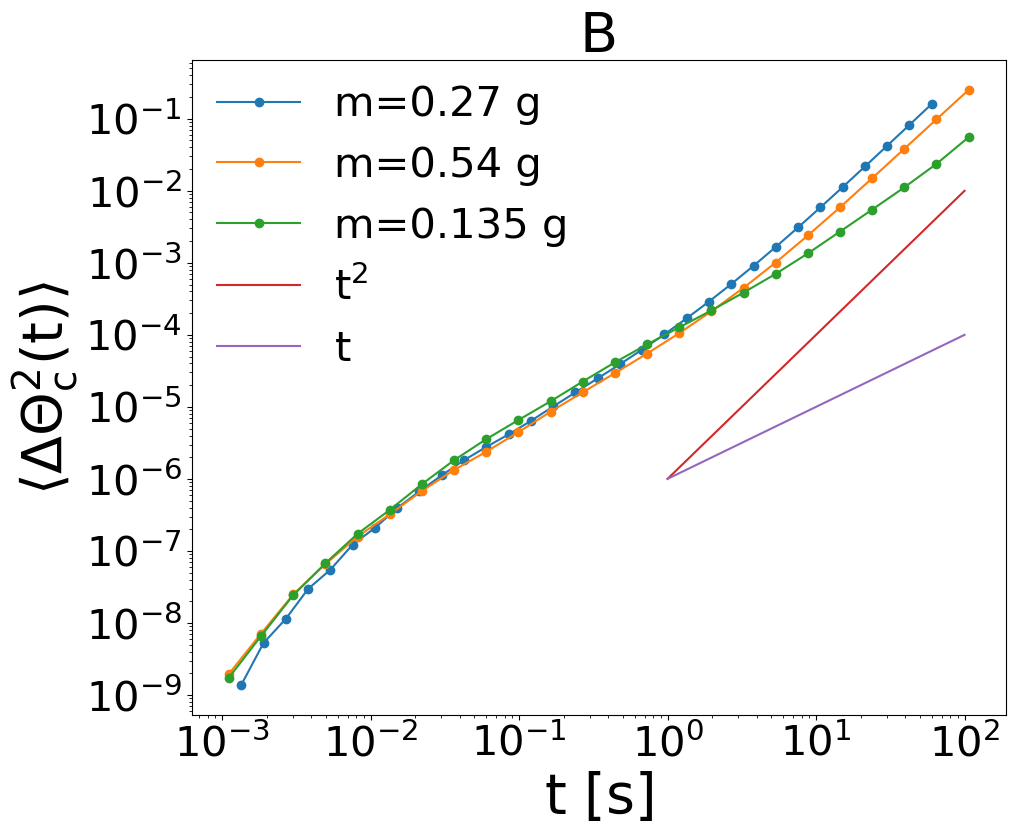}
\caption{Left: effect of inelasticity in the interaction among
  particles at constant granular temperature. The collective angular msd is shown
  for different values of the normal viscosity $\gamma_n$ (which also
  influences the tangential viscosity, see text). The average kinetic
  energy per particle is kept constant by appropriately tuning
  $\Gamma$. The three values of viscosity correspond to average
  restitution coefficients $e=0.90$, $e=0.99$ and $e=0.9999$
  respectively.  Right: effect - on the collective angular msd - of changing the mass of
  the particles at constant $\Gamma=39.8$. In both graphs we always
  have $N=2600$.  }
\label{dissmass}
\end{figure}

Inelasticity in our model mainly comes from the presence of viscous
interaction during contact, see Eqs.~\eqref{ForzaHM}, embodied in the
damping parameters $\gamma_n$ and $\gamma_t$, for normal and
tangential dissipation respectively. Here we change $\gamma_n$ keeping
constant the ratio $\gamma_t/\gamma_n$. The effect upon the mean
squared angular displacement of reducing the dissipation is shown in
Fig.~\ref{dissmass}A. We do not observe a clear monotonous behavior,
but a significant reduction of the superdiffusion phenomenon at
smaller dissipation can be excluded. This points at a possible
connection with long memory effects observed in elastic dense fluids,
for instance the so-called hydrodynamic long time tails in velocity
autocorrelations~\cite{brenig89}. It is however interesting to notice
that our system lives in three dimensions, where a hard sphere system
is expected to display velocity autocorrelations with tails decaying
as $\sim t^{-3/2}$ which are integrable and therefore exclude
superdiffusion. In fact superdiffusion in elastic three-dimensional
hard sphere systems has not been observed up to our
knowledge. Dimensionality, here, is of course not trivial since the
angular variables live in an effective single dimension. We also
recall that long time tails, of the same kind, are observed also in
systems of inelastic hard spheres~\cite{fiege2009}.

A more clear and relevant effect on msd, on the contrary, is seen when the
mass of the particles is reduced, see Fig.~\ref{dissmass}B. A system
with a mass $2$ times smaller appears to have weaker
superdiffusion. This seems coherent with our
interpretation of the superdiffusive regime as a consequence of large
inertia (together with coherence) in the system.

\end{document}